\documentclass[10pt,DIV16,twocolumn,numbers=noenddot]{scrartcl}

\usepackage[utf8]{inputenc}
\usepackage[T1]{fontenc}
\usepackage{lmodern}
\usepackage{amsmath,amssymb}
\usepackage[german]{authblk}
\usepackage[format=plain,labelfont={bf}]{caption}
\usepackage{listings}
\usepackage[shortcuts]{extdash} 
\usepackage{xcolor}
\usepackage{booktabs}
\usepackage{hyperref}
\usepackage{multirow}
\usepackage[sorting=none,citestyle=numeric-comp,eprint=false,isbn=false]{biblatex}
\setkomafont{disposition}{\bfseries}
\allowdisplaybreaks

\newcommand{\mytitle}{Comparison of methods for the calculation of the real dilogarithm regarding instruction-level parallelism}
\newcommand{\myauthor}{Alexander Voigt}
\newcommand{\Li}{\operatorname{Li}_2}
\newcommand{\secref}[1]{Section~\ref{#1}}
\newcommand{\tabref}[1]{\tablename~\ref{#1}}

\hypersetup{
  pdftitle   = {\mytitle},
  pdfauthor  = {\myauthor},
  colorlinks = true,
  linkcolor  = blue,
  citecolor  = blue,
  urlcolor   = blue,
  linkcolor  = blue
}

\title{\mytitle}
\author{\myauthor}
\affil{Fachbereich Energie und Biotechnologie, Hochschule Flensburg,\\ Kanzleistra{\ss}e 91--93, 24943 Flensburg, Germany}
\date{\today}

\begin{document}
\maketitle

\section*{Abstract}

We compare different methods for the computation of the real
dilogarithm regarding their ability for using instruction\-/level
parallelism when executed on appropriate CPUs.  As a result we present
an instruction\-/level\-/aware method and compare it to existing
implementations.

\section{Introduction}

The dilogarithm \cite{lewin} is a special function, which appears in
many physics applications, for example in the calculation of quantum
corrections in quantum field theory.  The investigation of new models
beyond the Standard Model of particle physics with large parameter
spaces require the numerical evaluation of the dilogarithm (and other
functions) billions of times.  Such investigations thus require a fast
evaluation of the dilogarithm.  In the past many time-efficient
algorithms have been developed (see e.g.\
\cite{koelbigDilog,ginsberg,morris,luke}) and many are implemented,
for example, in general-purpose physics and mathematics program
libraries \cite{root,gsl,cephes}.

Many of the publicly available implementations, e.g.\ \cite{root,gsl},
use formulations for the approximation of the dilogarithm with a
strictly sequential execution of floating\-/point operations.  Here
the development of modern CPUs with support for instruction\-/level
parallelism (ILP) opens the opportunity for a further reduction of the
run\-/time of the calculation of the dilogarithm.  To make use of ILP,
the algorithm to calculate the dilogarithm must be formulated
appropriately in order to allow for the simultaneous execution of
multiple floating\-/point operations.  In this paper a method for the
calculation of the real dilogarithm $\Li$ is presented, which makes
use of ILP, resulting in a reduced run\-/time compared to many
algorithms presented so far, when executed on appropriate CPUs.

This paper is organized as follows: In \secref{sec:toymodel} different
implementation strategies are discussed using a toy example function.
In \secref{sec:algorithm} an ILP-aware method for the numerical
calculation of the dilogarithm is presented.  A C implementation of
this method can be found in the appendix and in the arXiv submission
of this paper.

\section{Implementation strategies}
\label{sec:toymodel}

As a toy example we consider the real function $\ln(1+x)$ for $x>-1$,
which has the Taylor expansion
\begin{align}
  \ln(1+x) \approx x - \frac{x^2}{2} + \frac{x^3}{3} - \frac{x^4}{4} + \frac{x^5}{5} - \frac{x^6}{6} + \frac{x^7}{7}
  \label{eq:taylor}
\end{align}
around $x=0$ for $x\in(-1,1]$.  In Eq.~\eqref{eq:taylor} terms of
$O(x^8)$ and beyond have been omitted for brevity.  In the following
we will briefly show various possible implementations of this function
and discuss their performance implications.  For brevity we will
restrict ourselves to $x\in(-1,1]$.  In C the function in
Eq.~\eqref{eq:taylor} could be implemented very naively as follows:
\begin{lstlisting}[language=C]
double log1p_naive(double x) {
   return x - 1./2*x*x + 1./3*x*x*x
      - 1./4*x*x*x*x + 1./5*x*x*x*x*x
      - 1./6*x*x*x*x*x*x + 1./7*x*x*x*x*x*x*x;
}
\end{lstlisting}
This naive implementation requires 27 floating\-/point multiplications
and 6 floating\-/point additions.  Although all summands in this
implementation can in principle be calculated in parallel, the cost of
the many (redundant) floating\-/point multiplications is usually so
high that this formulation is in total very time\-/inefficient.

As is well known, the performance of this naive implementation can be
improved by applying Horner's scheme \cite{horner}, which reduces the
number of expensive floating\-/point multiplications to 7 and requires
6 additions:
\begin{align}
\begin{split}
  \ln(1+x) \approx{}& x \bigg\{1 + x \bigg[-\frac{1}{2} + x \bigg(\frac{1}{3} + x \bigg\{-\frac{1}{4} \\
  &+ x \bigg[\frac{1}{5} + x\bigg(-\frac{1}{6} + \frac{x}{7}\bigg)\bigg]\bigg\}\bigg)\bigg]\bigg\}.
\end{split}\label{eq:horner}%
\end{align}
A C implementation using Horner's scheme could read
\begin{lstlisting}[language=C]
double log1p_horner(double x) {
   return x*(1 + x*(-1./2 + x*(1./3
     + x*(-1./4 + x*(1./5
     + x*(-1./6 + 1./7*x))))));
}
\end{lstlisting}
Note that the formulation using Horner's scheme requires the
evaluation of the floating\-/point operations in strictly sequential
order.  However, the significant reduction of the number of
floating\-/point multiplications, compared to the naive implementation
from above, usually results in a shorter run\-/time.

There are several different possibilities to reduce the run\-/time of
Horner's method by rewriting the series expansion of $\ln(1+x)$ such
that more terms can be calculated in parallel, while the number of
floating\-/point operations is still kept small.  One possibility is
to split the r.h.s.\ of Eq.~\eqref{eq:horner} into two distinct sums,
one containing the even powers and the other containing the odd powers
of $x$,
\begin{align}
\begin{split}
  \ln(1+x) \approx{}& x \left\{1 + y \left[\frac{1}{3} + y
  \left(\frac{1}{5} + \frac{y}{7}\right)\right]\right\} \\
  &- y\left[\frac{1}{2} + y\left(\frac{1}{4} + \frac{y}{6}\right)\right],
\end{split}\label{eq:split}%
\end{align}
where $y=x^2$.  This formulation requires 8 floating\-/point
multiplications and 6 additions, i.e.\ one more multiplication than
Horner's scheme.  However, in Eq.~\eqref{eq:split} both sums can
be calculated in parallel, leading to a potential speed-up if the cost
of the additional multiplication is smaller than the gain by the
parallel execution.  A C implementation could read:
\begin{lstlisting}[language=C]
double log1p_split(double x) {
   const double y = x*x;
   return x*(1 + y*(1./3 + y*(1./5 + 1./7*y)))
      - y*(1./2 + y*(1./4 + 1./6*y));
}
\end{lstlisting}
Another possibility is to rewrite the r.h.s.\ of Eq.~\eqref{eq:taylor}
using Estrin's scheme \cite{estrin},
\begin{align}
\begin{split}
  \ln(1+x) \approx{}& x + y \left(\frac{x}{3}-\frac{1}{2}\right)
    \\ &+ z \left[\frac{x}{5} - \frac{1}{4}
      + y \left(\frac{x}{7}-\frac{1}{6}\right)
    \right],
\end{split}\label{eq:estrin}%
\end{align}
where $y=x^2$ and $z=y^2$.  This form also requires 8 floating\-/point
multiplications and 6 additions, as in Eq.~\eqref{eq:split}.  However,
in Estrin's form more terms can be executed in parallel, leading to a
potential further speed-up.  A C implementation using Estrin's scheme
could read:
\begin{lstlisting}[language=C]
double log1p_estrin(double x) {
   const double y = x*x;
   const double z = y*y;
   return x + y*(1./3*x - 1./2) +
      z*(1./5*x - 1./4 + y*(-1./6 + 1./7*x));
}
\end{lstlisting}

Another option to exploit ILP is to use a rational function
approximation, such as a Padé approximant \cite{pade}, where the
numerator and the denominator can be calculated in parallel.  A Padé
approximant of $\ln(1+x)$ at the same order as Eq.~\eqref{eq:taylor}
can be written as
\begin{align}
  \ln(1+x) &\approx x
  \frac{
    1 + \frac{51064}{40143} x + \frac{44320}{120429} x^2 + \frac{320}{40143} x^3
  }{
    1 + \frac{23712}{13381} x + \frac{12320}{13381} x^2 + \frac{5120}{40143} x^3
  }
  \label{eq:pade}
  \\
  &= x
  \frac{
    1 + \frac{51064}{40143} x + y \left(\frac{44320}{120429} + \frac{320}{40143} x\right)
  }{
    1 + \frac{23712}{13381} x + y \left(\frac{12320}{13381} + \frac{5120}{40143} x\right)
  },
  \label{eq:mixed}
\end{align}
where $y=x^2$.  In Eq.~\eqref{eq:mixed} the numerator and the
denominator have been re-written using Estrin's scheme to reduce the
number of floating\-/point multiplications compared to
Eq.~\eqref{eq:pade}, while still allowing for some degree of ILP at
the same time.  A C implementation of Eq.~\eqref{eq:mixed} could read:
\begin{lstlisting}[language=C]
double log1p_mixed(double x) {
   const double y = x*x;
   const double num = 1 + 51064./40143*x +
      y*(44320./120429 + 320./40143*x);
   const double den = 1 + 23712./13381*x +
      y*(12320./13381 + 5120./40143*x);
   return x*num/den;
}
\end{lstlisting}
Note that the value $y=x^2$ can be re-used in the calculation of both
the numerator and the denominator.  A computation of $y^2$ is not
necessary.  This ``mixed'' implementation requires 8 floating\-/point
multiplications, 1 floating\-/point division and 6 additions.  Note
also that a rational function approximant such as Eq.~\eqref{eq:mixed}
naturally allows for ILP, as the numerator and denominator can be
computed in parallel.  This is in particular beneficial for long
numerator/denominator polynomials, where the cost of the
floating\-/point division is smaller than the gain by the parallel
computation of the terms in the numerator and denominator.  In order
to further reduce the error and/or the number of terms in the
approximant, a more ``optimized'' minimax rational function
approximant may be used instead of a Taylor expansion or Padé
approximant when approximating a function over a larger interval.

\section{ILP-aware approximant of the real dilogarithm}
\label{sec:algorithm}

In the following an ILP-aware implementation of the real dilogarithm
with double precision is presented, which allows for the simultaneous
execution of multiple floating\-/point operations.  The real
dilogarithm is defined as
\begin{align}
  \Li(x) &= -\int_0^x \frac{\ln(1-t)}{t} \, \textnormal{d}t, &
  &x<1.
\end{align}
For $|x|<1$ the real dilogarithm has the series representation
\begin{align}
  \Li(x) = \sum_{k=1}^\infty \frac{x^k}{k^2}.
  \label{eq:series}
\end{align}
Further series representations exist, for example the accelerated
series
\begin{align}
  \Li(x) = \sum_{k=0}^\infty \frac{B_k}{(k+1)!} [-\ln(1-x)]^{k+1},
\end{align}
where $B_k$ are the Bernoulli numbers.  A series representation in
terms of Chebyshev polynomials of the first kind $T_n(x)$ is given in
Ref.~\cite{luke},
\begin{align}
  \Li(x) = \sum_{k=0}^\infty a_n T_n(2x-1),
  \label{eq:cheby}
\end{align}
which is used for example in the ROOT program library \cite{root}.
The r.h.s.\ of Eqs.~\eqref{eq:series}--\eqref{eq:cheby} can be summed,
for example, using Horner's scheme or Clenshaw's algorithm
\cite{clenshaw}, respectively.  These summation techniques are,
however, purely sequential and thus do not make use of ILP.  In the
following an approximant of the real dilogarithm is presented, which
makes use of ILP, using a rational function approximation in
combination with Estrin's scheme, similar to the approach used in
Eq.~\eqref{eq:mixed}.

We define the real dilogarithm $\Li(x)$ for all $x\in\mathbb{R}$ as
the real part of its complex continuation.  The real dilogarithm can
be mapped onto the interval $[0,1/2]$ by using the following
identities \cite{lewin}:
\begin{align}
  \Li(x) ={}& \ln(1-x)\left[\frac{1}{2}\ln(1-x) - \ln(-x)\right] - \frac{\pi^2}{6}
              \notag \\ &+\Li\left(\frac{1}{1-x}\right), \quad x\leq -1, \\
  \Li(x) ={}& -\Li\left(\frac{x}{x-1}\right) - \frac{1}{2}\ln^2(1-x),
              \notag \\ & -1<x<0, \\
  \Li(x) ={}& -\Li(1-x) + \frac{\pi^2}{6} - \ln(x) \ln(1-x),
              \notag \\ &\frac{1}{2}<x<1, \\
  \Li(x) ={}& \frac{\pi^2}{6} - \ln(x)\left[\ln\left(1-\frac{1}{x}\right) + \frac{1}{2}\ln(x)\right]
              \notag \\ &+ \Li\left(1-\frac{1}{x}\right), \quad 1<x\leq 2, \\
  \Li(x) ={}& -\Li\left(\frac{1}{x}\right) + \frac{\pi^2}{3} - \frac{1}{2} \ln^2(x), \quad x>2,
\end{align}
and by using the special values $\Li(-1)=-\pi^2/12$, $\Li(0)=0$,
$\Li(1/2)=\pi^2/12-\ln^2(2)/2$, $\Li(1)=\pi^2/6$ and $\Li(2)=\pi^2/4$.
Within the interval $[0,1/2]$ the real dilogarithm has the series
expansion \eqref{eq:series}, which can be written as
\begin{align}
  \Li(x) = x \sum_{k=0}^\infty \frac{x^{k}}{(k+1)^2}.
  \label{eq:series2}
\end{align}
We approximate the sum on the r.h.s.\ of Eq.~\eqref{eq:series2} in the
form of a rational minimax approximant on the interval $[0,1/2]$ using
the \texttt{MiniMaxApproximation} from Wolfram/Mathematica
\cite{mathematica},
\begin{align}
  \sum_{k=0}^\infty \frac{x^{k}}{(k+1)^2} \approx
  \frac{\sum_{k=0}^5 p_kx^k}{\sum_{k=0}^6 q_kx^k}.
  \label{eq:minimax}
\end{align}
The coefficients $p_k$ and $q_k$ are listed in \tabref{tab:coeffs}.
The approximant on the r.h.s.\ of Eq.~\eqref{eq:minimax} has a maximum
relative error of around $5\cdot 10^{-17}$.  This error is small
enough so that the approximant Eq.~\eqref{eq:minimax} can be used for
the evaluation of $\Li(x)$ with double precision.
\begin{table}[t]
  \centering
  \caption{Coefficients of the numerator and denominator polynomials
    for the minimax approximant \eqref{eq:minimax}.}
  \begin{tabular}{lr}
    \toprule
    $p_0$ & $ 0.9999999999999999502\cdot 10^{+0}$ \\
    $p_1$ & $-2.6883926818565423430\cdot 10^{+0}$ \\
    $p_2$ & $ 2.6477222699473109692\cdot 10^{+0}$ \\
    $p_3$ & $-1.1538559607887416355\cdot 10^{+0}$ \\
    $p_4$ & $ 2.0886077795020607837\cdot 10^{-1}$ \\
    $p_5$ & $-1.0859777134152463084\cdot 10^{-2}$ \\
    $q_0$ & $ 1.0000000000000000000\cdot 10^{+0}$ \\
    $q_1$ & $-2.9383926818565635485\cdot 10^{+0}$ \\
    $q_2$ & $ 3.2712093293018635389\cdot 10^{+0}$ \\
    $q_3$ & $-1.7076702173954289421\cdot 10^{+0}$ \\
    $q_4$ & $ 4.1596017228400603836\cdot 10^{-1}$ \\
    $q_5$ & $-3.9801343754084482956\cdot 10^{-2}$ \\
    $q_6$ & $ 8.2743668974466659035\cdot 10^{-4}$ \\
    \bottomrule
  \end{tabular}
  \label{tab:coeffs}
\end{table}
Note that the approximant on the r.h.s.\ of Eq.~\eqref{eq:minimax} is
very similar to the approximant presented in Ref.~\cite{morris} (DILOG
0011), although with slightly different coefficients.  In order to
allow for the use of ILP in the evaluation of the numerator and the
denominator on the r.h.s.\ of Eq.~\eqref{eq:minimax} we use Estrin's
scheme and write
\begin{align}
  \Li(x) &\approx x\frac{P(x)}{Q(x)}, & &0\leq x \leq \frac{1}{2},
  \label{eq:final}
\end{align}
with
\begin{align}
  P(x) &= p_0 + p_1 x + (p_2 + p_3 x) y + (p_4 + p_5 x) z, \\
  Q(x) &= q_0 + q_1 x + (q_2 + q_3 x) y + (q_4 + q_5 x + q_6 y) z,
\end{align}
where $y=x^2$ and $z=y^2$.  A C implementation of Eq.~\eqref{eq:final}
can be found in the appendix.

\section{Run\-/time comparison}

In \tabref{tab:runtime} the run\-/time of different C implementations
for the real dilogarithm is compared to the implementation given in
the appendix on two different CPU architectures.  On each architecture
the implementations are compiled with \texttt{g++} with optimization
level \texttt{-O2}.  The shown run\-/time denotes the total wall clock
time for the calculation of $\Li(x)$ for $10^6$ uniformly distributed
random values $x\in[0,1/2]$.  The choice of this interval ensures that
the measurement is only sensitive to the formulation of the
approximant, but not to the transformation onto the interval
$[0,1/2]$.  Note that in applications where
$x\in\mathbb{R}\backslash [0,1/2]$, the transformation onto the
interval $[0,1/2]$ can have a significant impact on the run\-/time.

\begin{table}[tb]
  \centering
  \caption{Run\-/time in seconds for the calculation of $\Li(x)$ for
    different implementations on different CPU architectures with
    different compilers.}
  \begin{tabular}{lcc}
    \toprule
    \multirow{2}{*}{Implementation} & i7-4700MQ & i7-5600U \\
    & \texttt{g++} 9.3.0 & \texttt{g++} 10.2.1 \\
    \midrule
    Eq.~\eqref{eq:final} (this paper) & $0.0076$ & $0.0058$ \\
    Morris (DILOG 0011) \cite{morris} & $0.0115$ & $0.0085$ \\
    Cephes \cite{cephes}              & $0.0136$ & $0.0119$ \\
    Algorithm 327 \cite{koelbigDilog} & $0.0229$ & $0.0172$ \\
    Chebyshev \cite{luke,root}        & $0.0718$ & $0.0515$ \\
    Algorithm 490 \cite{ginsberg}     & $0.1057$ & $0.0576$ \\
    GSL \cite{gsl}                    & $0.2091$ & $0.0898$ \\
    \bottomrule
  \end{tabular}
  \label{tab:runtime}
\end{table}%

We find that the implementation presented in this paper and the
approach presented in Ref.~\cite{morris} (DILOG 0011) are fastest
among the ones shown in the table.  Both approaches use a rational
function approximant with the same number of terms.  However, the
approximant from Ref.~\cite{morris} has been implemented (for
comparison) to use Horner's scheme to evaluate the numerator and the
denominator.  Thus, the run\-/time difference of around $30\%$ between
the two stems only from the fact that the implementation presented in
this paper uses Estrin's scheme to evaluate the numerator and the
denominator and as such can make use of ILP, while in the
implementation of Ref.~\cite{morris} the numerator and the denominator
are evaluated sequentially.

The third fastest implementation is provided by Cephes \cite{cephes},
which implements a rational function approximation for Spence's
function $\operatorname{Sp}(x)=\Li(1-x)$.  The implementation provided
by Cephes is by a factor two slower than the implementation presented
in this paper.  This longer run\-/time has several reasons: For one,
Cephes uses a rational function approximation with more terms in the
numerator and denominator.  Furthermore, Cephes evaluates the
numerator and denominator polynomials using Horner's scheme, which is
purely sequential.%

Algorithm 327 \cite{koelbigDilog}, Algorithm 490 \cite{ginsberg} and
the implementation found in the GNU Scientific Library (GSL)
\cite{gsl} use series expansions, which are summed using Horner's
scheme.  These algorithms thus do not make use of ILP, which is one
reason for their longer run\-/time.  Furthermore, Algorithm 490 and
most series expansions in the GSL use expensive floating\-/point
divisions inside loops, which may be another reason for the increased
run\-/time.

The representation in terms of Chebyshev polynomials \cite{luke} is
implemented in the ROOT program library \cite{root}.  In this
implementation the polynomials are summed using Clenshaw's algorithm,
which is also purely sequential and slightly more complex than
Horner's scheme.  This is a reason why the implementation in
\cite{root} has a larger run\-/time than the rational function
approximants discussed above.

\section{Summary}

In this paper different approaches for the numerical calculation of
the real dilogarithm have been compared w.r.t.\ their potential for
the use of instruction\-/level parallelism (ILP).  We found that a
rational minimax approximant in combination with Estrin's scheme to
evaluate the numerator/denominator polynomials provides both a
sufficient numerical precision and a high degree of ILP, potentially
leading to a reduced run\-/time when executed on a appropriate CPUs.
A corresponding approximant and a C implementation have been presented
and compared to several existing implementations w.r.t.\ their
run\-/time.  The C implementation of the presented approximant can be
found in the appendix.  C++, Fortran, Julia and Rust implementations
can be found in \cite{polylogarithm,PolyLog.jl,polylog}, respectively.

\appendix

\section{Implementation of the real dilogarithm}

\begin{lstlisting}[language=C]
#include <math.h>

double li2(double x)
{
   const double PI = 3.1415926535897932;
   const double P[] = {
      0.9999999999999999502e+0,
     -2.6883926818565423430e+0,
      2.6477222699473109692e+0,
     -1.1538559607887416355e+0,
      2.0886077795020607837e-1,
     -1.0859777134152463084e-2
   };
   const double Q[] = {
      1.0000000000000000000e+0,
     -2.9383926818565635485e+0,
      3.2712093293018635389e+0,
     -1.7076702173954289421e+0,
      4.1596017228400603836e-1,
     -3.9801343754084482956e-2,
      8.2743668974466659035e-4
   };

   double y = 0, r = 0, s = 1;

   if (x < -1) {
      const double l = log(1 - x);
      y = 1/(1 - x);
      r = -PI*PI/6 + l*(0.5*l - log(-x));
      s = 1;
   } else if (x == -1) {
      return -PI*PI/12;
   } else if (x < 0) {
      const double l = log1p(-x);
      y = x/(x - 1);
      r = -0.5*l*l;
      s = -1;
   } else if (x == 0) {
      return 0;
   } else if (x < 0.5) {
      y = x;
      r = 0;
      s = 1;
   } else if (x < 1) {
      y = 1 - x;
      r = PI*PI/6 - log(x)*log(y);
      s = -1;
   } else if (x == 1) {
      return PI*PI/6;
   } else if (x < 2) {
      const double l = log(x);
      y = 1 - 1/x;
      r = PI*PI/6 - l*(log(y) + 0.5*l);
      s = 1;
   } else {
      const double l = log(x);
      y = 1/x;
      r = PI*PI/3 - 0.5*l*l;
      s = -1;
   }

   const double y2 = y*y;
   const double y4 = y2*y2;
   const double p = P[0] + y*P[1]
      + y2*(P[2] + y*P[3])
      + y4*(P[4] + y*P[5]);
   const double q = Q[0] + y*Q[1]
      + y2*(Q[2] + y*Q[3])
      + y4*(Q[4] + y*Q[5] + y2*Q[6]);

   return r + s*y*p/q;
}
\end{lstlisting}

\printbibliography

\end{document}